# Competing exchanges and spin-phonon coupling in $Eu_{1-x}R_xMnO_3$ (R=Y,Lu)

D. A. Mota[1], Y. Romaguera Barcelay[1], P. B. Tavares[2], M.R. Chaves[1], A. Almeida[1], J. Oliveira[1], W. S. Ferreira[3], J. Agostinho Moreira[1,a)]

[1]IFIMUP and IN-Institute of Nanoscience and Nanotechnology, Departamento de Física e Astronomia da Faculdade de Ciências, Universidade do Porto, Rua do Campo Alegre, 687, 4169-007 Porto, Portugal.

[2]Centro de Química – Vila Real. Universidade de Trás-os-Montes e Alto Douro. Apartado 1013, 5001-801. Vila Real. Portugal.

[3]GRUMA - Grupo de Magnetoeletricidade, Departamento de Física, Centro de Educação, Ciências Exatas e Naturais, Universidade Estadual do Maranhão. Cidade Universitária Paulo VI, s/n - Tirirical, CEP: 65055-970. São Luís, Maranhão. Brasil.

**ABSTRACT**

This work is focused on the phase diagrams and physical properties of Y-doped and Lu-doped EuMnO3. The differences in the corresponding phase boundaries in the (x,T) phase diagram could be overcome by considering a scaling of the $Y^{3+}$ and $Lu^{3+}$ concentrations to the tolerance factor. This outcome evidences that the tolerance factor is in fact a more reliable representative of the lattice deformation induced by doping. The normalization of the phase boundaries using the tolerance factor corroborates previous theoretical outcomes regarding the key role of competitive FM and AFM exchanges in determining the phase diagrams of manganite perovskites. Though, significant differences in the nature and number of phases at low temperatures and concentrations could not be explained by just considering the normalization to the tolerance factor. The vertical phase boundary observed just for Lu-doped EuMnO3, close to 10%Lu, is understood considering a low temperature Peierls-type spin-phonon coupling, which stabilizes the AFM4 phase in Lu-doped EuMnO3.

a) Electronic mail: jamoreir@fc.up.pt

# I. INTRODUCTION

In recent years a particular attention is focused on systems that present coupled magnetic and ferroelectric properties.[1-4] Among these systems, the study of rare-earth manganites ($RMnO_3$) revealed new properties and rich phase diagrams.[4-8] Some modulated structures are coupled to ferroelectricity, which show interesting theoretical and experimental results.[9-14]

In rare-earth manganites, the coupling between magnetism and ferroelectricity can be well described by Dzyaloshinskii-Moriya interaction (DM) by assuming a coupling between electrical spin and orbital electrical motion, where atomic displacements are not formally included.[9-14] In fact C. D. Hu,[11] taking as starting point the original works of T. Morya[15] and I. Dzyaloshinskii[16] emphasized the consequences of hybridization of p orbitals of oxygen atoms in those systems, enhancing the super-exchange antiferromagnetic (AFM) interactions against the ferroelectric ones, which provides the driving force for the emergence of magnetically-induced, ferroelectric ground states.

The deviation from an ideal perovskite towards an orthorhombic distorted structure of rear-earth manganites is associated with two major geometrical mechanisms. One mechanism, also known as the $GdFeO_3$-type distortion, involves the tilt of $MnO_6$ octahedra, occurring alternately in opposite directions along the crystallographic *c*-direction, if P*bnm* notation is used.[17,18] This distortion, characterized by the magnitude of the bond angle Mn-O-Mn, connecting $Mn^{3+}$ and octahedral apical oxygen ions depends on the size of the R ion ($r_R$), and the smaller $r_R$ is, the smaller is the tilt angle.[14,17-19] The other mechanism is the distortion of $MnO_6$ octahedral imposed by the Jahn-Teller effect associated with the $Mn^{3+}$ ion.[19]

Furthermore, associated with $Mn^{3+}$ spins, several types of exchanges have been considered, which were then introduced in the microscopic model developed by M. Mochizuki *et al*[14] to discuss phase diagrams and the role of spin-phonon coupling in those materials. These exchanges can be summarized as follows: (i) nearest-neighbour ferromagnetic (FM) exchanges $J_{ab}$, along x and y, next nearest-neighbour AFM exchanges $J_a$ and $J_b$, along a and b, and AFM exchanges $J_c$ along *c*.[14]

Under the assumption that $J_b$ can be induced and enhanced by the $GdFeO_3$-type distortion, the ($T$-$r_R$) diagrams for $RMnO_3$, and the ($T$-$r_R$) diagrams for solid solutions were successfully confirmed by ($T$-$J_b$) phase diagrams, obtained by varying the magnitude of $J_b$. In order to study their magnetoelectric phase diagrams, avoiding other magnetic interactions other than those coming from the $Mn^{3+}$ ion, non-magnetic rare-earth ions manganites and their non-magnetic ion-doped solid solutions have been considered, as it is the case–study the $Eu_{1-x}Y_xMnO3$ system.[20-22] The great advantage of this solid solution is that by increasing dopant-concentration only the effect of geometrical mechanisms and thus $J_b$ are expected to influence the phase diagram and spin-phonon coupling, enabling to determine their role in defining the nature and number of different phases. Along with the aforementioned solid solution, $Eu_{1-x}Lu_xMnO_3$ is also very interesting to be addressed to, since besides keeping the non-magnetic nature of the dopant ion, its radius being smaller than the yttrium one enables to also figure out the effect of the radius magnitude and thus of the balance between competing FM and AFM exchanges in tailoring both phase diagrams and physical properties in rare-earth manganites.[23]

Though the magnetoelectric phase diagrams of both derivates were previously published, some intriguing differences in both their phase diagrams and physical properties still remain to be understood. It is the aim of this work to discuss these differences thoroughly, within the scope of both GdFeO$_3$-type lattice distortion and balance between competitive FM and AFM exchanges, as well as by considering a spin-phonon coupling mechanism.

## II. PHASE-DIAGRAMS

The ($x$,T) phase diagrams of the systems $Eu_{1-x}Y_xMnO_3$, $0 \leq x \leq 0.5$ and $Eu_{1-x}Lu_xMnO_3$, with $0 \leq x \leq 0.3$ are displayed in Figure 1, which are well described in the current.[23,24]

Before addressing to the main issues of the phase diagrams, it is worthwhile to note that the corresponding maximal composition is not the same. This is because the compounds for higher concentrations of either lutetium or yttrium become multiphasic, and thus cannot be considered to trace their phase-diagrams. In order to have a more straight analysis, we have shown the two phase-diagrams in Figure 1 using a common concentration scale.

Figure 1 shows that both systems undergo a paramagnetic–antiferromagnetic phase transition (PM-AFM1) at the Néel temperature ($T_N$) phase-boundary. $T_N$ exhibits the same concentration dependence for both systems and decreases very slowly with $x$. The AFM1 phase has been considered as a non-ferroelectric, magnetic, sinusoidal collinear modulated phase.

Besides AFM-1, a cycloidal spin arrangement modulated AFM-2 phase is also observed for a Y-concentration between 15-50% and a Lu-concentration between 10-30%. The AFM-2 phase shown in Figure 1, allows for ferroelectricity, according to the inverse Dzyaloshinskii-Moriya (DM) model.[11,15,16] Besides other minor differences, three deserve to be emphasized. One lies on the low-concentration value marking the onset of AFM-2 phase, as it is seen from Figures

1(a) and (b), which actually emerges for a lower concentration of $Lu^{3+}$. The second one refers to the low temperature limiting phase boundary of the AFM-2 phase. While for the Lu-doped system it is almost a straight line (Figure 1(a)), in the other system it starts to bend to low temperatures after a 30% concentration of yttrium is reached (Figure 1(b)). The third one is most pertinent as it lies on the nature of the lowest temperature phases. In order to simplify the discussion let us rename these phases shown in Figure 1, as follow: (i) As the cA-AFM, WFM phase is a week ferromagnetic, canted AFM phase, just as the AFM-3 in the phase diagram of the Y-doped system, let us call both AFM-3; (ii) Instead, the AFM-3, paraelectric phase is an ordinary AFM phase, and will be call further on as AFM-4. It is then clear that at low temperature the two systems have different concentration-dependent phases, specified by the dashed vertical phase line shown in Figure 1(a), though missing in Figure 1(b). While the Lu-doped system undergoes a phase transition from AFM-3 to AFM-4, close to 10% Lu, characterized by spin canting suppression, in the Y-doped system spin canting is preserved at least up to below 40% Y.

### III. RESULTS

#### a. Lattice distortions and phase-diagrams

In this section, lattice alterations induced by the substitution of the $Eu^{3+}$ ion by the non-magnetic $Y^{3+}$ and $Lu^{3+}$ ions will be explored, trying to comprehend the aforementioned differences based on mechanisms associated with lattice distortions.

Deviations from the ideal perovskite structure can be determined by using the Goldschmidt tolerance factor $t$, defined as follows:[25]

$$t = \frac{r_A + r_o}{\sqrt{2}(r_b + r_o)},$$

where $r_A$ and $r_B$ are the ionic radii of the ions in A site, coordinating number (CN) 8, and B site (CN 6) respectively, and $r_O$ the ionic radius of $O^{2-}$ for CN 6.[26] Early X-ray diffraction and Raman studies in both systems revealed that the substitution of $Eu^{3+}$ ion by $Y^{3+}$ or $Lu^{3+}$ is associated with lattice distortions characterized by a decrease in the tolerance factor and a reduction of the unit cell volume. Moreover, those studies evidenced also that both reductions correlate with Mn-O1-Mn bond angle decrease. Contrarily, no correlation was found with the octahedral Mn-O bond lengths that remain essentially unchanged. So, another way to evaluate the deviation from an ideal structure may be obtained from the Mn-O1-Mn angle, which in ideal conditions is 180°.

Figure 2(a) shows the relation between the dopant concentration $x$ and the tolerance factor $t$, calculated from equation (1), taking:

$$r_a = (1-x)r_{Eu} + xr_d$$

where $r_{Eu}$ and $r_d$ are the ionic radii of $Eu^{3+}$ and the dopant ($Y^{3+}$ or $Lu^{3+}$), respectively, for CN=8.[26]

As it is confirmed from Figure 2, $t$ is in fact a decreasing function of $x$. For the Lu-doped system the $t(x)$ slope is lower than for the yttrium one, which is not surprising as the $Lu^{3+}$ radius is smaller than the $Y^{3+}$ one.[26] From Figure 2, we conclude that the dopant concentration is not the best parameter to scale distortions. In order to sort out a unique lattice-distortion scaling parameter for both systems, we chose to analyze the behaviour of the Mn-O1-Mn bond angle as a function of tolerance factor, seen in Figure 3(a).

In the range of studied compositions there is approximately a linear relation between those values, which evidences that both are equivalent parameters for characterizing distortions from the ideal perovskite structure. Similar conclusion is achieved from the analysis of Figure 3(b), which shows the $t$-dependence of the Ag(4) tilt mode frequency.[27] It is worthwhile to note that the $B_{2g}$ symmetric stretching mode frequency (see Ref. 27), shown in the inset of Figure 3(b), is practically independent of $t$, also evidencing similar frequency values obtained for both systems. These results are in good agreement with lattice distortions induced by $EuMnO_3$-doping, which mainly involve the tilting of the $MnO_6$ octahedra.[20,23] The scaling of both Mn-O1-Mn bond angle and Ag(4) tilt mode frequency with the tolerance factor suggests that doped-mediated lattice distortions play a major role in both systems. Thus, in the following both phase diagrams we will be rescaled as a function of the tolerance factor, instead of the concentration as it was shown in Figure 1.

Figure 4 shows the phase diagrams of the $Eu_{1-x}Y_xMnO_3$ and $Eu_{1-x}Lu_xMnO_3$ systems, where the horizontal axis is normalized to the tolerance factor.

The more awesome result is the fact that taking the tolerance factor as a scaling parameter, similar phase boundaries lines are observed in both diagrams. From Figure 4 we also reckon that for $x_Y$=0.2 and $x_{Lu}$=0.1 are very closely located in phase diagram. This is also the case for the compositions $Eu_{0.5}Y_{0.5}MnO_3$ and $Eu_{0.7}Lu_{0.3}MnO_3$ where it can be observed the same phase sequence, namely they both exhibit a re-entrant ferroelectric phase. These results emphasize the importance of geometrical parameters to determine the critical temperatures of the different phase transitions. In fact, the tilting of MnO$_6$ octahedra, which is associated with the tolerance factor, determines in large extent the similitude of both phase diagrams.

However, the tilting of the octahedra is not enough to explain some significant differences between both phase diagrams, in particular the low temperature, low concentration phase transition from AFM-3 to AFM-4 occurring just in Eu$_{1-x}$Lu$_x$MnO$_3$. This phase transition is marked in Figure 4 by a red dashed column. Other differences, regarding the behaviour of their physical properties, will be touched on further below.

### b. Magnetic ground states at low temperatures

In the following we address to the low temperature, low concentration range of the phase diagrams shown in Figure 4. Though in this range the compositions Eu$_{0.8}$Y$_{0.2}$MnO$_3$ and Eu$_{0.9}$Lu$_{0.1}$MnO$_3$ are located close together, their phase sequence looks different: the former stabilizes in a week FM, canted AFM-3 phase and the latter in a non-canted AFM-4 one.

Figure 5 shows the induced magnetization of the aforementioned compounds as a function of temperature, measured in zero-field cooling (ZFC) and field-cooling (FC) runs, using a 40 Oe magnetic field (see experimental details in Ref. 23). Before describing the results of Figure 5, it is worthwhile to note that special care has to be taken in interpreting induced magnetization curves, when manganite perovskites are being studied. In these systems, the driving forces, based on the competition between FM and AFM usually yield magnetic disorder. Thus, measuring magnetization using an auxiliary magnetic field, though arbitrary small, additional contributions may emerge due to magnetic disorder, leading to misleading outcomes regarding the actual magnetic ground states.

Let us first touch on the ZFC curves shown in Figure 5 for both compounds. In the 20% Y-doped compound a weak ferromagnetic component exists below T$_N$, but only below 20 K is an intrinsic contribution associated with the weak FM, canted AFM-3 ground state, as it was confirmed by M(H) curves.[28] Contrarily, in the temperature range from 20K to T$_N$, where an ordinary AFM is stable, the magnetization contribution is just being induced by the auxiliary magnetic field. The T-dependence of M(T) for the 10% Lu-doped system is totally different, yielding a non- canted AFM-4 phase below 20K, which is revealed from the decreasing shape of the M(T) curve with temperature decrease.[23] Though, it is important to stress that the FC curve for 10% Lu-doped system is similar to the FC curve of the 20% Y-doped one. In fact, the former should be placed in such a position in the (*t*,T) phase diagram, where magnetic disorder is enough high to enable an induced magnetization in the whole temperature range below to $T_N$. Despite their closeness in the phase diagram of Figure 4, the nature of the low temperature magnetic ground states for both compounds is in fact different.

In order to get further information regarding the microscopic mechanism stabilizing the low temperature phases, we have performed a detailed lattice dynamics study through Raman spectroscopy (see experimental details in Ref. 20). According to the spin-phonon coupling models, one should expect detectable changes in the phonon frequencies on entering the magnetic phases, reflecting the phonon renormalization, proportional to the spin-spin correlation function for the nearest $Mn^{3+}$ spins.[29-31] Aiming at searching for a spin-phonon coupling in Y- and Lu-doped $EuMnO_3$, we have monitored the temperature dependence of the wavenumber of the Ag(4)-lattice bending mode associated with the tilt of the $MnO_6$ octahedra for both $Eu_{0.9}Lu_{0.1}nO_3$, and $Eu_{0.8}Y_{0.2}MnO_3$, which is shown in Figure 6.

Though the temperature evolution of the wavenumber in the paramagnetic phase is already distinctly different, which has been ordinarily assigned to magnetic disorder-mediated fluctuations, the most striking issue is the way Ag(4)-mode wavenumber changes below to $T_N$. Whilst in $Eu_{0.9}Lu_{0.1}nO_3$ an increase of the wavenumber is observed with decreasing temperature, in $Eu_{0.8}Y_{0.2}MnO_3$ a decrease of the wavenumber occurs.

In order to figure out the mechanisms, which are subjacent to the observed shifts, the theoretical model expressed by equation (1) will be used.[29-31] This model states that in the case ferromagnetic and antiferromagnetic competitive interactions exist, $\omega - \omega_o$ wavenumber shift can be given as:[31]

$$\omega - \omega_o \propto -R_1 \langle \vec{S}_i | \vec{S}_j \rangle + R_2 \langle \vec{S}_i | \vec{S}_k \rangle, \qquad (1)$$

where $\omega_o$ is the frequency in the absence of spin-phonon coupling, $R_1$ and $R_2$ are spin dependent force constants of the lattice vibrations deduced as the squared derivatives of the exchange integrals with the respect to the phonon displacement.[31] Whilst $R_1$ is associated with nearest ferromagnetic exchange $J_{ab}$, $R_2$ reflects the antiferromagnetic next-nearest neighbour one $J_b$.[31] This model predicts negative or positive frequency shifts depending on the relative strength between the ferromagnetic and antiferromagnetic exchange interactions, associated with the mode being considered.

As it has been assumed in current literature, we consider that the spin correlation functions of the nearest neighbours and the next-nearest neighbours have almost the same temperature dependence, and thus, we take the same correlation functions $\langle \vec{S}_i | \vec{S}_j \rangle$ and $\langle \vec{S}_i | \vec{S}_k \rangle$. Moreover, we also take constant values for $R_1$ and $R_2$ for the same vibration mode. Thus, Equation (1) can be written as equation (2):[20]

$$\omega - \omega_o \propto (R_2 - R_1)\langle \vec{S}_i | \vec{S}_j \rangle. \qquad (2)$$

Since for $Eu_{0.8}Y_{0.2}MnO_3$ the Raman shift is negative, corresponding to a negative difference $R_2 - R_1$, FM $J_{ab}$ exchange overcome the AFM $J_b$ ones, yielding the stabilization of the weak FM, AFM-3 phase for this compound. On the contrary, for $Eu_{0.9}Lu_{0.1}MnO_3$ the difference $R_2 - R_1$ is positive, due to the positive shift of the Ag(4) wavenumber, favouring the AFM $J_b$ exchanges against the FM $J_{ab}$ ones, and thus stabilizing at low temperatures the non-canted AFM4 phase. This interpretation is fully in agreement with the existence of the vertical phase boundary for Lu-doped EuMnO3, which separates at low temperatures the AFM3 from the AFM4 phase, as it is shown in Figure 4.

The reason why this transition occurs only in Lu-doped EuMnO3 cannot rely on the tilting of $MnO_6$ octahedra as for $x_{Lu}=0.1$ and $x_Y=0.2$ their values deviates just 0,1%. Lutetium has to have an additional effect on the balance between FM $J_b$, and AFM $J_{ab}$ exchanges in order to stabilize the AFM4 phase at low temperatures.

Before going to details, let us recall the mechanism referred to Masahito *et al*[14] concerning the magnitude of the nearest-neighbour FM exchange $J_{ab}$ on the Mn-O-Mn bonds along the pseudocubic x and y axes. Since its magnitude depends sensitively on the in-plane Mn-O-Mn angle, a Peierls-type spin-phonon coupling were considered ($J'_{ab}\delta_{ij}$) reflecting the contribution to $J_{ab}$ exchange, from the shift $\delta_{ij}$ of the in-plane oxygen ions relatively to its orthorhombic position, associated to magnetic ordering at low temperatures.[14,32]

Assuming that Lutetium induces a sufficiently high positive shift $\delta_{ij}$, a decrease of the magnitude of FM $J_{ab}$ exchange will necessary occur, since to the negative room temperature value of Jab a positive low temperature term will be added.[14] The decrease of nearest neighbour FM $J_{ab}$ exchange against the next-nearest neighbour AFM $J_b$ exchange will be then the driving force to stabilize the AFM4 phase at low temperatures in Lu-doped EuMnO3. From this mechanism, two main issues stand out. On one hand, the strengthening of the FM $J_{ab}$ exchanges against the AFM $J_b$ ones, obtained from considering the low temperature Peierls-type spin-phonon contribution,[14] is in perfect agreement with the outcome obtained by applying the model expressed in equation (1) to the Raman data. On the second hand, the Lu-induced shift $\delta_{ij}$ is in good agreement with the wavenumber shift $\omega - \omega_o$ of the Ag(4) mode in

Lu-doped EuMnO$_3$ (Figure 6(a)), since the decrease of the in-plane Mn-O-Mn angle, reflecting a higher strain level, yields an increase of wavenumber. It would be very interesting to study A-site lattice distortion, as it could give a confirmation of the interpretation presented to above.

It is worthwhile to stress that specific differences between both systems do not confine just to the phase diagrams themselves, but also extend to the shape and amplitude of the temperature dependence of their physical properties. Though both (*t*,T) phase diagrams were traced from the anomalies observed in the temperature dependence of the dielectric, polar, magnetic, and magnetoelectric properties as well as the specific heat, their shapes and magnitudes are different.[23,24,28] These differences stem from a variety of parameters, whose nature have been largely referred to in earlier published works: spin-phonon and spin-lattice coupling, spin-orbit and spin-exchange interactions, and frustration-mediated spiral spin orders.[11-16] One interesting example is to observe the temperature dependence of the temperature/composition dependence of the spontaneous polarization for both systems shown in Figures 14 and 1 of Refs. 23 and 24, respectively. Though the emergence of polarization is expected from the inverse DM interaction in the spiral-spin incommensurate phases, their amplitudes and shapes mirror, as it is expected from equation (1) of Ref. 14, the effect of spin-orbit and spin-exchange interaction, as well as of spiral-spin magnetic ordering. These mechanisms are in fact expected to be dependent on the nature and concentration of the dopant ion, as it is the case of Y- and Lu doped EuMnO$_3$ systems.

## IV. CONCLUSIONS

This work is focused on the phase diagrams and physical properties of Y-doped and Lu-doped EuMnO3, in order to figure out which are the relevant driving mechanisms through a detailed analysis of their similarities and differences. Excluding the low temperature, low concentration part of the phase diagrams, the nature and number of phases are similar. The differences in the corresponding phase boundaries could be overcome by considering a scaling of the concentration to the tolerance factor. This outcome evidences that the tolerance factor is in fact a more reliable representative of the lattice deformation induced by doping, and thus of the balance between competitive FM and AFM exchanges. The normalization of the phase boundaries using the tolerance factor corroborates previous theoretical outcomes from M. Mochizuki et al. regarding the key role of competitive FM and AFM exchanges in determining the phase diagrams of manganite perovskites.

Though, differences in the nature and number of phases at low temperatures and concentrations cannot be explained by just considering the normalization to the tolerance factor.

The vertical phase boundary observed just for Lu-doped EuMnO3 close to 10%Lu, could be understood if a low temperature Peierls-type spin-phonon contribution is considered. This mechanism, which strengths the AFM $J_b$ exchanges against the FM $J_{ab}$ ones, acts as the driving force that stabilizes the AFM4 phase in Lu-doped EuMnO3.

**Acknowledgments**

This work was supported by the Fundação para a Ciência e Tecnologia and COMPETE/QREN/EU, through the project PTDC/CTM/099415/2008.

**References**

[1] N. A. Hill, J. Phys. Chem. B **104**, 6694 (2000).

[2] W. Eerenstein, N. D. Mathur, and J. F. Scott, Nature (London) **442**, 759 (2006).

[3] R. E. Cohen, Nature **358**, 136 (1992).

[4] T. Kimura, and Y. Tokura, J. Phys.: Condens. Matter **20**, 434204 (2008).

[5] T. Kimura, S. Ishihara, H. Shintani, T. Arima, K. T. Takahashi, K. Ishizaka, and Y. Tokura, Phys. Rev. B 68, 060403(R) (2003).

[6] T. Kimura, G. Lawes, T. Goto, Y. Tokura, A. P. Ramirez, Phys. Rev. B 71, 224425 (2005).

[7] J. S. Lee, N. Kida, Y. Yamasaki, R. Shimano, and Y. Tokura, Phys. Rev. B 80, 134409 (2009).

[8] Y. Yamasaki, H. Sagayama, T. Goto, M. Matsuura, K. Hirota, T. Arima, and Y. Tokura, Phys. Rev. Lett. 98, 147204 (2007).

[9] I. A. Sergienko and E. Dagotto, Phys. Rev. B 73, 094434 (2006).

[10] M. Mostovoy, Phys. Rev. Lett. 96, 067601 (2006).

[11] C. D. Hu, Phys. Rev. B 77, 174418 (2008).

[12] Masahito Mochizuki and Nobuo Furukawa, Phys. Rev. B 80, 134416 (2009).

[13] Masahito Mochizuki, Nobuo Furukawa, and Naoto Nagaosa, Phys. Rev. Lett. 105, 037205 (2010).

[14]Masahito Mochizuki, Nobuo Furukawa, and Naoto Nagaosa, Phys. Rev. B 84, 144409 (2011).

[15]T. Moriya, Phys. Rev 120, 91 (1960).

[16]I. Dzyaloshinsky, J. Phys. Chem. Solids 4, 241 (1958).

[17]A. M. Glazer, Acta Cryst. B28, 3384 (1972).

[18]A. M. Glazer, Acta Cryst. A31, 756 (1975).

[19]T. Goto, T. Kimura, G. Lawes, A. P. Ramirez, and Y. Tokura, Phys. Rev. Lett. 92, 257201 (2004).

[20]J. Agostinho Moreira, A. Almeida, W. S. Ferreira, J. P. Araújo, A. M. Pereira, M. R. Chaves, J. Kreisel, S. M. F. Vilela and P. B. Tavares, Phys. Rev. B 81, 054447 (2010).

[21]J. Hemberger, F. Schrettle, A. Pimenov, P. Lunkenheimer, V. Yu. Ivanov, A. A. Mukhin, A. M. Balbashov, and A. Loidl, Phys. Rev. B 75, 035118 (2007).

[22]Y. Yamasaki, S. Miyasaka, T. Goto, H. Sagayama, T. Arima, and Y. Tokura, Phys. Rev. B 76, 184418 (2007).

[23]J. Oliveira, J. Agostinho Moreira, A. Almeida, M. R. Chaves, J. M. M. da Silva, J. B. Oliveira, M. A. Sá, P. B. Tavares, R. Ranjith, and W. Prellier, Phys. Rev. B 84, 094414 (2011).

[24]J. Agostinho Moreira, A. Almeida, W.S. Ferreira, M.R. Chaves, J.B. Oliveira, J.M. Machado da Silva, M.A. Sá, S.M.F. Vilela, P.B. Tavares, Solid State Communications 151, 368 (2011).

[25]R. H. Mitchel, Perovskites, Modern and Ancient. Almaz Press Inc. Canada. 2002.

[26]R. D. Shannon, Acta Cryst. A32, 751 (1976).

[27]M. N. Iliev, M. V. Abrashev, J. Laverdière, S. Jandl, M. M. Gospodinov, Y.-Q. Wang, and Y.-Y. Sun, Phys. Rev. B 73, 064302 (2006).

[28]J. Agostinho Moreira, A. Almeida, W. S. Ferreira, M. R. Chaves, J. P. Araújo, A. M. Pereira, S. M. F Vilela and P. B. Tavares, J. Phys.: Condens. Matter 22, 125901 (2010).

[29]D. J. Lockood and M. G. Cottam, J. Appl. Phys. 64, 5876 (1988).

[30]W. Baltensperger and J. S. Helman, Helv. Phys. Acta 41, 668 (1968).

[31]K. Wakamura, and T. Arai, J. Appl. Phys. 63, 5824 (1988).

[32]Y. Lépine, Phys. Rev. B 28, 2659 (1983).

**Figure captions**

Figure 1. The ($x$,T) phase diagrams of the systems (a) $Eu_{1-x}Lu_xMnO_3$, 0≤x≤0.3 and (b) $Eu_{1-x}Y_xMnO_3$, with 0≤x≤0.5.

Figure 2. Tolerance factor as a function of the dopant concentration.

Figure 3. Mn-O1-Mn bond angle (a) and Ag(4) Raman active tilt mode (b) as a function of the tolerance factor for $Eu_{1-x}Y_xMnO_3$ and $Eu_{1-x}Lu_xMnO_3$. Inset: B2g Raman active symmetric stretching mode as a function of the tolerance factor. All values were determined at room temperature.

Figure 4. Superposed phase diagrams of $Eu_{1-x}Y_xMnO_3$ and $Eu_{1-x}Lu_xMnO_3$ systems, normalized to the tolerance factor for the corresponding contents of $Y^{3+}$ (down scale) or $Lu^{3+}$ (up scale). Dashed vertical column, AFM-3 and AFM-4 in red belongs exclusively to the $Eu_{1-x}Lu_xMnO_3$ phase diagram.

Figure 5. Magnetization of the $Eu_{0.8}Y_{0.2}MnO_3$ and $Eu_{0.9}Lu_{0.1}MnO_3$ a function of temperature, measured in zero-field cooling (ZFC) and field-cooling (FC) runs, using a 40 Oe magnetic field.

Figure 6: Temperature dependence of the Ag-mode wavenumber for $Eu_{0.9}Lu_{0.1}nO_3$ (a) and $Eu_{0.8}Y_{0.2}MnO_3$ (b).

**Figures**

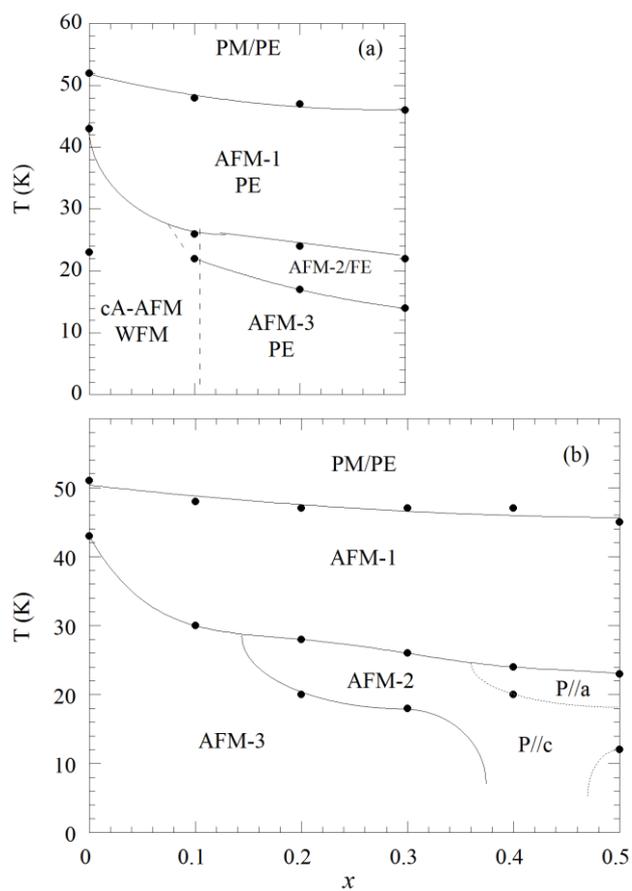

Figure 1

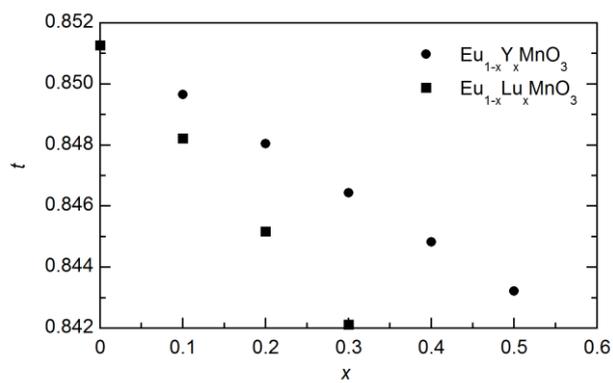

Figure 2

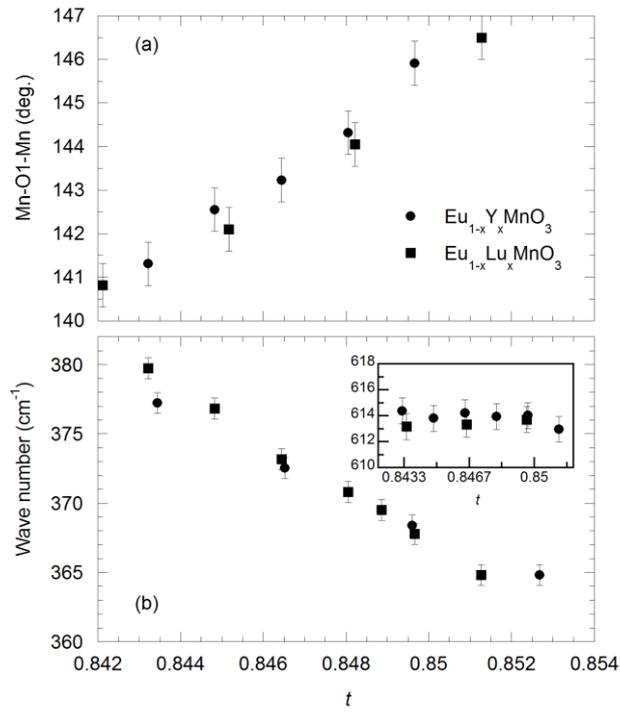

Figure 3

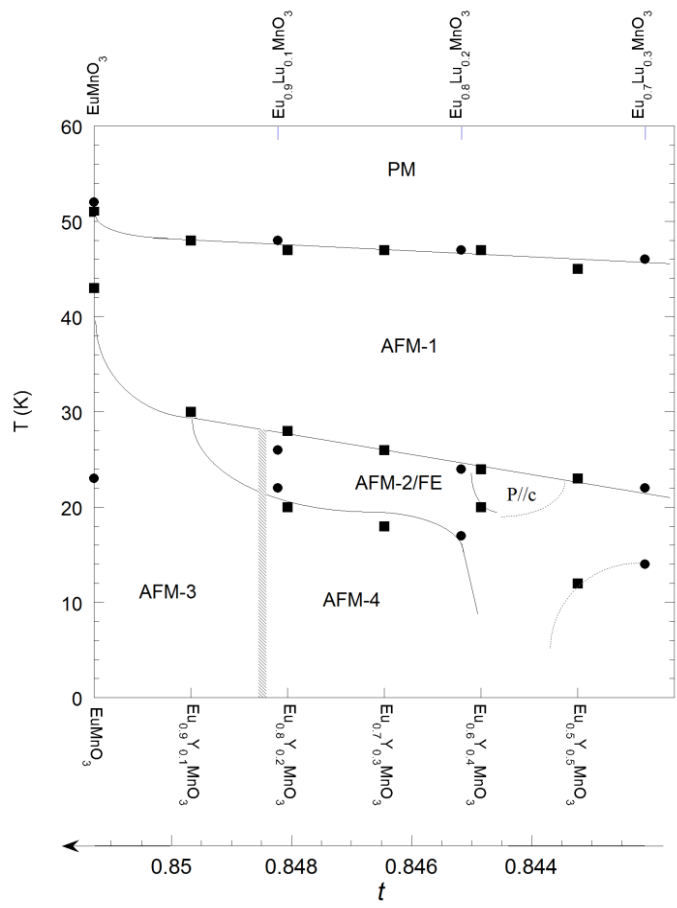

Figure 4

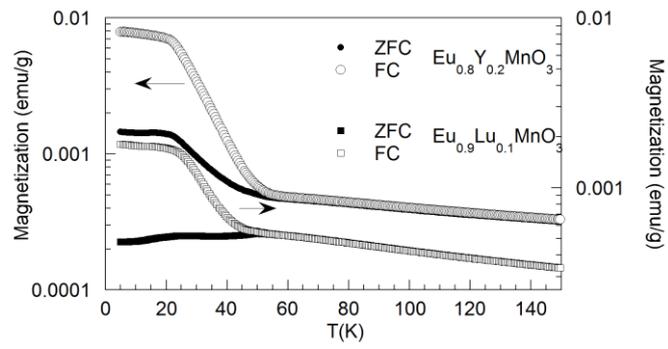

Figure 5

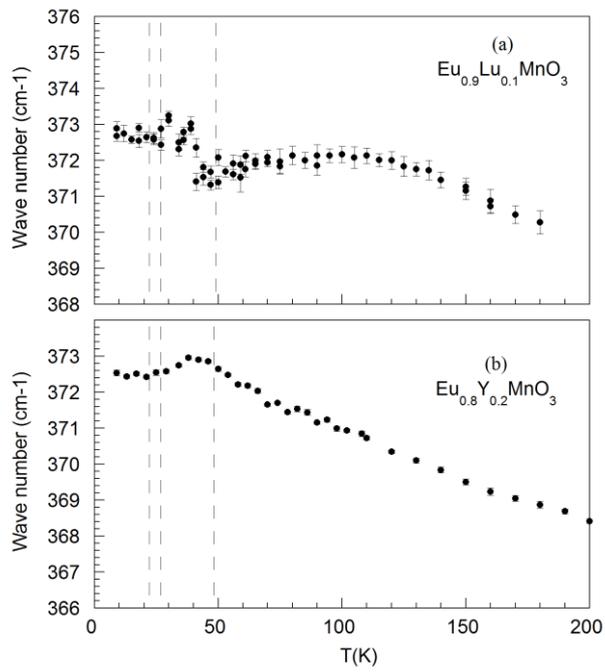

Figure 6